\documentclass[aps,pre,twocolumn,superscriptaddress,showpacs,amsmath,amssymb]{revtex4-1}

\usepackage{graphicx}% Include figure files

\begin{document}

\title{Living bacteria rheology: population growth, aggregation patterns and cooperative behaviour under different shear flows}

\author{P. Patr\'{\i}cio}
\affiliation{ISEL,
Rua Conselheiro Em\'{\i}dio Navarro 1, P-1959-007 Lisboa, Portugal}
\affiliation{CEDOC, Faculdade de Ci\^encias M\'edicas, Universidade Nova de Lisboa, 1169-056, Lisboa, Portugal}

\author{P. L. Almeida}
\affiliation{ISEL,
Rua Conselheiro Em\'{\i}dio Navarro 1, P-1959-007 Lisboa, Portugal}
\affiliation{CENIMAT/I3N, Faculdade Ci\^encias e Tecnologia, Universidade Nova de Lisboa, 2829-516 Caparica, Portugal}

\author{R. Portela}
\affiliation{Centro de Recursos Microbiol\'ogicos, Faculdade de Ci\^encias e Tecnologia, Universidade Nova de Lisboa, 2829-516 Caparica, Portugal}

\author{R. G. Sobral}
\affiliation{Centro de Recursos Microbiol\'ogicos, Faculdade de Ci\^encias e Tecnologia, Universidade Nova de Lisboa, 2829-516 Caparica, Portugal}

\author{I. R. Grilo}
\affiliation{Laborat\'orio de Gen\'etica Molecular, ITQB, Universidade Nova de Lisboa, 2780 Oeiras, Portugal}

\author{T. Cidade}
\affiliation{CENIMAT/I3N, Faculdade Ci\^encias e Tecnologia, Universidade Nova de Lisboa, 2829-516 Caparica, Portugal}
\affiliation{Departamento de Ci\^encia dos Materiais, Faculdade de Ci\^encias e Tecnologia, Universidade Nova de Lisboa, 2829-516 Caparica, Portugal}

\author{C. R. Leal}
\email{cleal@adf.isel.pt}
\affiliation{ISEL,
Rua Conselheiro Em\'{\i}dio Navarro 1, P-1959-007 Lisboa, Portugal}
\affiliation{CENIMAT/I3N, Faculdade Ci\^encias e Tecnologia, Universidade Nova de Lisboa, 2829-516 Caparica, Portugal}

\date{\today}

\begin{abstract}

The activity of growing living bacteria was investigated using real-time and {\it in situ} rheology -- in stationary and oscillatory shear.
Two different strains of the human pathogen {\it Staphylococcus aureus} -- strain COL and its isogenic cell wall autolysis mutant -- were considered in this work.
For low bacteria density, strain COL forms small clusters, while the mutant, presenting deficient cell separation, forms irregular larger aggregates.
In the early stages of growth, when subjected to a stationary shear, the viscosity of both strains increases with the population of cells.
As the bacteria reach the \textit{exponential phase} of growth, the viscosity of the two strains follow different and rich behaviours,
with no counterpart in the optical density or in the population's colony forming units measurements.
While the viscosity of strain COL keeps increasing during the \textit{exponential phase} and returns close to its initial value for the \textit{late phase} of growth, where the population stabilizes,
the viscosity of the mutant strain decreases steeply, still in the \textit{exponential phase}, remains constant for some time and increases again, reaching a constant plateau at a maximum value for the \textit{late phase} of growth.
These complex viscoelastic behaviours, which were observed to be shear stress dependent, are a consequence of two coupled effects:
the cell density continuous increase and its changing interacting properties. The viscous and elastic moduli of strain COL, obtained with oscillatory shear,
exhibit power-law behaviours whose exponent are dependent on the bacteria growth stage. The viscous and elastic moduli of the mutant have complex behaviours,
emerging from the different relaxation times that are associated with the large molecules of the medium and the self-organized structures of bacteria. These behaviours reflect nevertheless the  bacteria growth stage.

\end{abstract}

\maketitle

\section{Introduction}
\label{sec1}

The rheology or mechanical characterization of living systems, such as a bacterial culture, is a complex task and difficult to address.
The mechanical response of these systems will firstly depend on the type of bacteria that is analyzed (species and strain).
It will depend on the particular environment chosen to propagate the bacteria. It will depend also on the external physical stimuli imposed to the system.
However, the most challenging feature of these biological systems in what regards rheology is certainly the fact that they are dynamic,
changing their behaviour over time and space.

The primary physiological processes of a bacterial cell include, among others, cell grow and division, cell separation and cell-cell interaction. All these processes rely on the constant \textit{de novo} synthesis of the cell´s building blocks, including proteins, lipids and polysaccharides.
Furthermore, bacterial cells are able to detect the global density of the population (a mechanism called quorum sensing),
or other external stimuli such as an applied shear stress, acquiring a complex cooperative behaviour.

In recent years, increasing interest has been devoted to the mechanics of living cells \cite{Verdier2009,Mofrad2009,Huber2013},
with particular emphasis on the rheology of the cell cytoplasm \cite{Moeendarbary2013,Zhou2013}.
Although initially thought to be an aqueous environment, the cell internal milieu is today known to be comparable in its rheology to soft materials including polymeric solutions, gels, foams, etc.
However, due to their dynamic physiology, cells present important distinctive features which differentiate them from these familiar inert materials:
at rest, molecular fluctuations within their network are not dominated by thermal fluctuations, but rather by the ongoing release and consumption of energy
in the form of chemically activated small molecules such as adenosine triphosphate -- ATP \cite{Bursac2005};
for small imposed strains, the viscoelastic properties of the cells present scaling laws which identify them as soft glassy materials \cite{Fabry2001,Rogers2008,Wilhelm2008};
for larger deformations, cells may undergo a transition from a solid-like to a fluid-like state \cite{Trepat2007}.

Beyond the study of single cells, anomalous fluctuations were also detected in active bacterial suspensions \cite{Chen2007}.
Here, the diffusion of tracer particles in low-density bacterial cultures was characterized in detail, in well defined and short periods of time.
But probably because this study considered only low density cultures, the important changes occurring during growth as a consequence of cell's cooperative or social behaviour were not reported \cite{Ben-Jacob2008,Nadell2013}.

In general, bacteria present two main distinct types of growth, the planktonic growth, a suspension of dispersed bacterial cells which occurs in liquid environment and biofilm growth, a microbial community that adheres to a solid surface and is surrounded by a bacterially produced extracellular matrix. Biofilms are prevalent in natural environments and also in industrial and hospital settings. In the latter scenario, one of the major players and leading cause of hospital-acquired infections is {\it Staphylococcus aureus}, an opportunistic pathogenic bacterium famous for its virulence and capacity to accumulate antibiotic resistance traits.

Due to its important implications for public health and the environment, the rheology of bacterial biofilms has gained considerable attention during the last decade. Particular focus has been given to the viscoelasticity of cell attachment and detachment, under the influence of antibiotics and flows with different shear rates \cite{Klapper2002,Mascari2003,Rupp2005}. In these studies, the analysis of cells detachment was performed only in small time slots, at specific biofilm growth stages. More recently, particle-tracking microrheology was used to study communal lifecycles of bacteria, from growth to starvation, exploring possible cooperative behaviours emerging from quorum-sensing mechanisms \cite{Rogers2008}. By tracking cell displacements, the compliances of the biofilms were measured, showing power-law rheology with exponents that changed with incubation time.

While research efforts have focused on the study of biofilms, only scarce knowledge exists on the structure of planktonic populations and their capacity to aggregate. This is especially relevant for the study of pathogenic bacteria, as nonattached aggregates may be responsible for bacterial spreading in many clinical scenarios of persistent and chronic infections.

To characterize the dynamics and mechanical properties of planktonic populations, we have recently monitored the growth of {\it Staphylococcus aureus} strain COL \cite{Portela2013}. Real time and {\it in situ} rheology revealed a rich viscoelastic behaviour as a consequence of bacterial activity, namely, of the multiplication of cells and of density-dependent aggregation events.

We are presently extending our study to differentiate the population viscoelastic behaviour and aggregation patterns of two {\it S. aureus} strains, COL and an isogenic autolysin mutant, with different cell separation phenotypes.
In this work, we use RUSAL9, a transposition mutant strain impaired in the process of daughter cell separation, leading to the
formation of large cell clusters and a rougher cell surface.
Furthermore, and in order to understand how the bacterial population dynamics depends on environmental mechanical stimuli, we followed bacterial growth under different shear flow conditions.

This article is organised as follows: in Section \ref{sec2} we describe the bacterial strains considered in this study, and the various experimental techniques used to characterize them: bacteria propagation, optical density measurements, optical imaging and rheology.
In Section \ref{sec3} we analyze the development of the bacterial cultures, establishing their different distinct growth phases.
We characterize in detail the aggregating structures that are formed by each strain, and interpret the viscosity growth curves
obtained for stationary flow with different shear rates. We also analyze the viscoelastic behaviour of both strains under oscillatory flow, at different growth time points. We present our conclusions in Section \ref{sec4}.

\section{Experimental / Methods}
\label{sec2}

\subsection{Bacterial strains and growth conditions}

Two related strains of {\it Staphylococcus aureus} MRSA -- strain COL \cite{Gill2005} and its mutant derivative of \textit{atl} gene strain RUSAL9 \cite{Oshida1992} -- were used. Cultures were grown at 37 $^\circ$C with aeration in rich medium, tryptic soy broth (TSB, Difco Laboratories, Detroit, USA). Over-night grown li\-quid cultures were used to re-inoculate fresh medium at an initial $\text{OD}_{620\text{nm}}$ of 0.005, for rheological characterization.

\subsection{Optical density and Plating}

To monitor bacterial growth, we measured the optical density (620 nm) at discrete time intervals, resorting to a spetophotometer Ultrospec 2100 pro. In parallel, we also determined the population's colony forming units (cfus/ml), which provides an estimate of the viable cells, by plating serial dilutions of the bacterial cultures on tryptic soy agar (TSA, Difco). The plates were incubated for 48 h at 37 $^\circ$C, and the colonies were counted.

\subsection{Optical microscopy and Image analysis}

We characterized the size and number of cell aggregates during growth with optical microscopy, using a Leica DMR microscope with a Leica DFC320 camera and Leica IM500 Image software V1.20. For each aliquot, at specific growth moments, 250, 350, 410 and 550 min, a calibrated volume of sample of 10$\mu$L was collected and observed -- 10 photographs were taken randomly. From these images, we evaluated the cell distribution during growth, namely the number of clusters \textit{vs} the number of cells per cluster and the average number of cells per cluster \textit{vs} time.

\subsection{Rheology}

Rheological measurements were performed in a controlled stress rotational rheomenter Bohlin Gemini HR$^\text{nano}$. A steel plate/plate geometry, with diameter 40 mm and 2000 $\mu$m gap (to ensure a good signal), was used for the measurements of the viscosity growth curve, at a constant shear rate, at 37 $^\circ$C (to allow optimal bacterial growth). Measurements were performed at 5, 10 and 20~s$^{-1}$ (10 s$^{-1}$ mimics rpm incubator conditions). A steel cone/plate geometry, with diameter 40 mm, an angle of 2$^\circ$ and
70 $\mu$m gap, was used to perform oscillatory measurements. We measured the elastic and viscous moduli ($G'$ and $G''$) \textit{vs} the angular frequency $\omega$ in the linear regime, imposing 10\% of strain. Assays were performed at 20 $^\circ$C, to delay bacterial growth during measurements. A solvent trap was used in all measurements to avoid evaporation.

\section{Results and Discussion}
\label{sec3}

\subsection{Role played by cell aggregation during growth -- strain COL \textit{vs} strain RUSAL9}

\subsubsection{Growth phases and aggregation patterns}

\begin{figure}[htp]
\begin{center}
\includegraphics[scale=0.3]{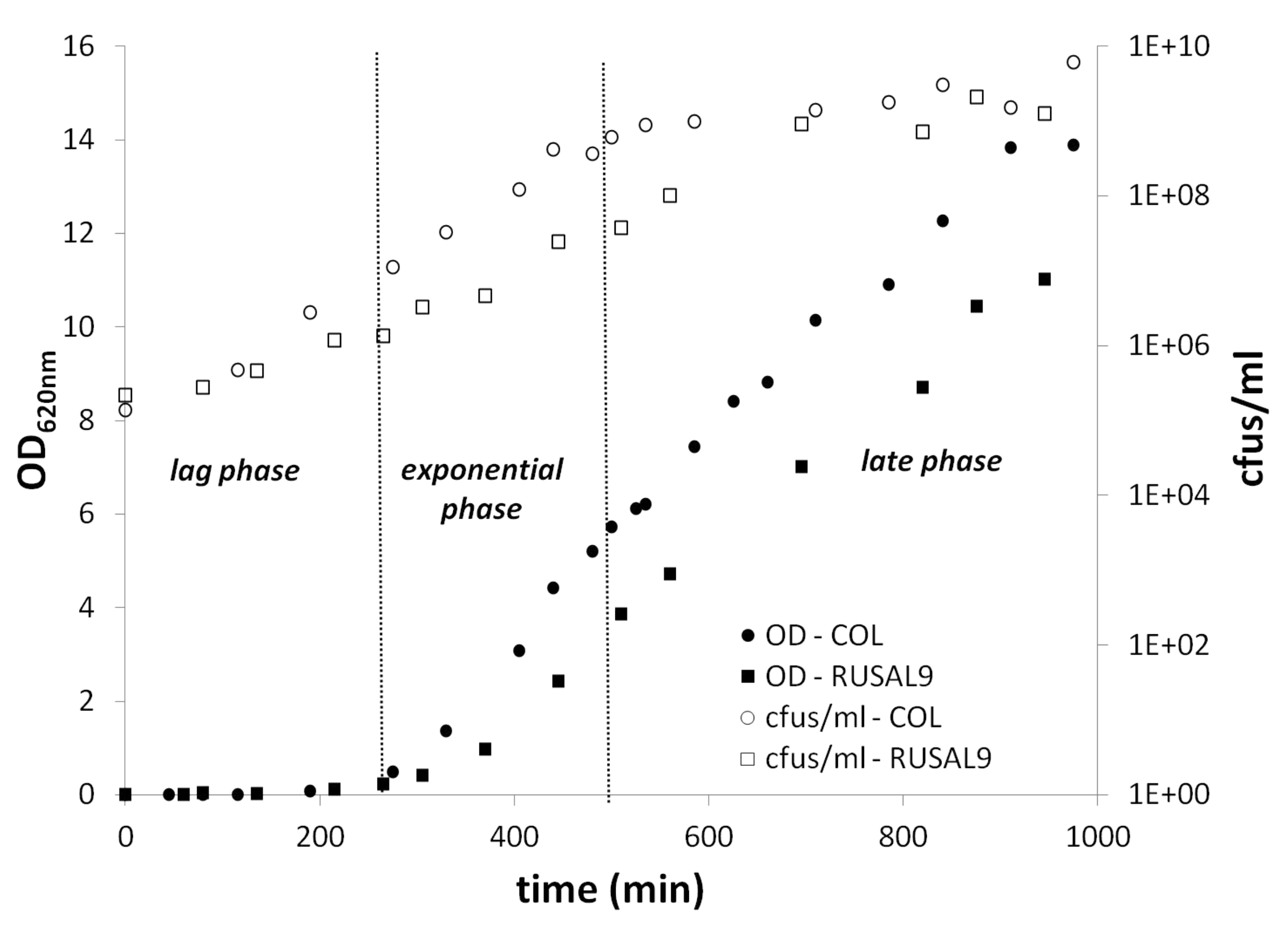}
\caption{\textit{S. aureus} cultures -- strain COL (circles) and strain RUSAL9 (squares) characterized by optical densities ($\text{OD}_{620\text{nm}}$) (full symbols) and population's colony-forming units (cfus/ml) (open symbols) during growth; dashed lines separate distinct growth phases: lag, exponential and late phases. All measurements were performed at 37 $^\circ$C.}
\label{fig1}
\end{center}
\end{figure}

Bacterial growth of \textit{S. aureus} cultures -- strain COL and strain RUSAL9 -- were monitored by measuring the optical density ($\text{OD}_{620\text{nm}}$) at discrete time intervals, in parallel with population's colony-forming units (cfus/ml). Representative curves are included in Fi\-gure \ref{fig1}. From these measurements it is possible to identify three different growth phases: \textit{lag phase}, \textit{exponential phase} and \textit{late phase}. No significant differences are observed when comparing strain COL and strain RUSAL9.

\begin{figure}[htp]
\begin{center}
\begin{tabular}{cc}
\includegraphics[scale=0.35]{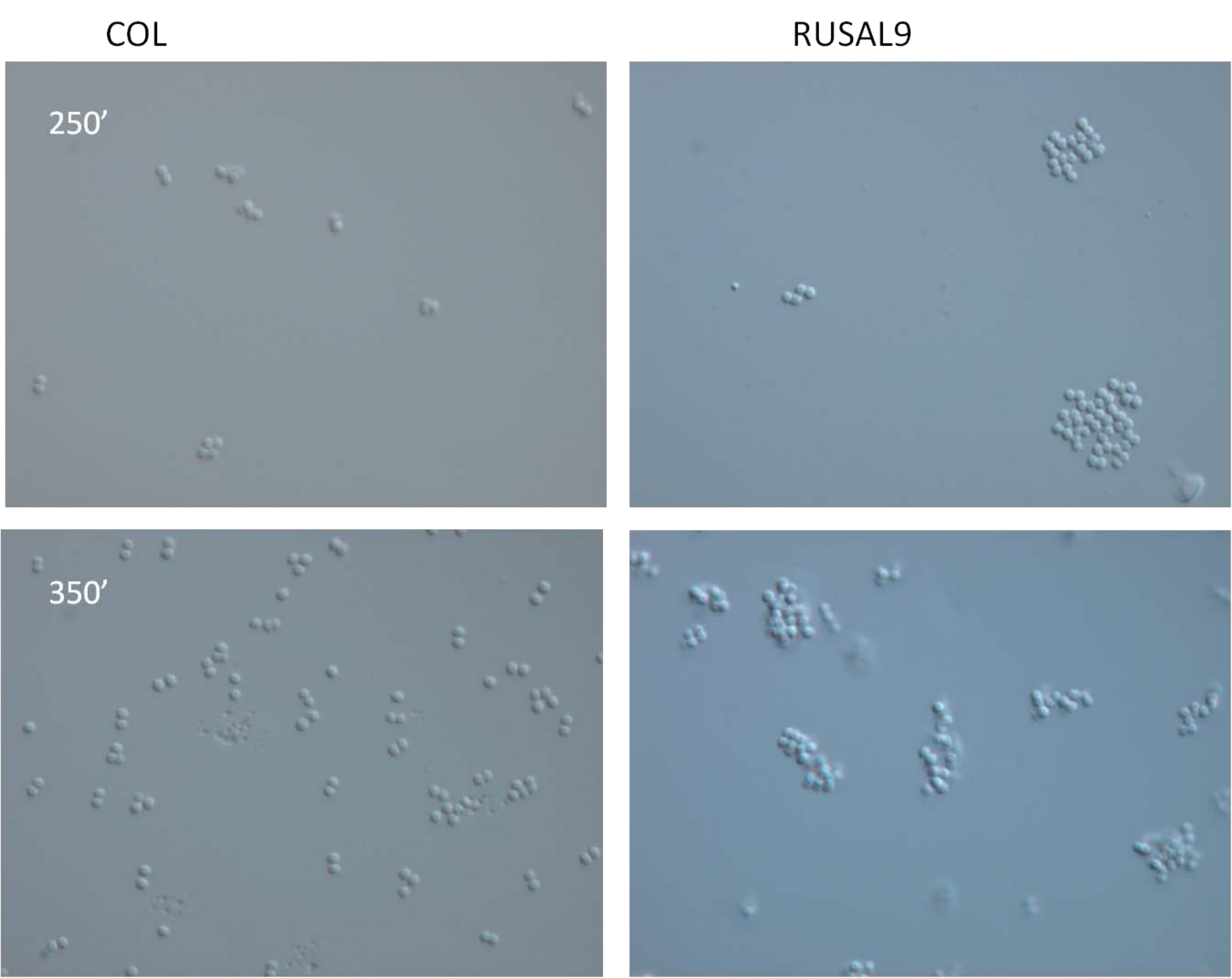}\\
\includegraphics[scale=0.35]{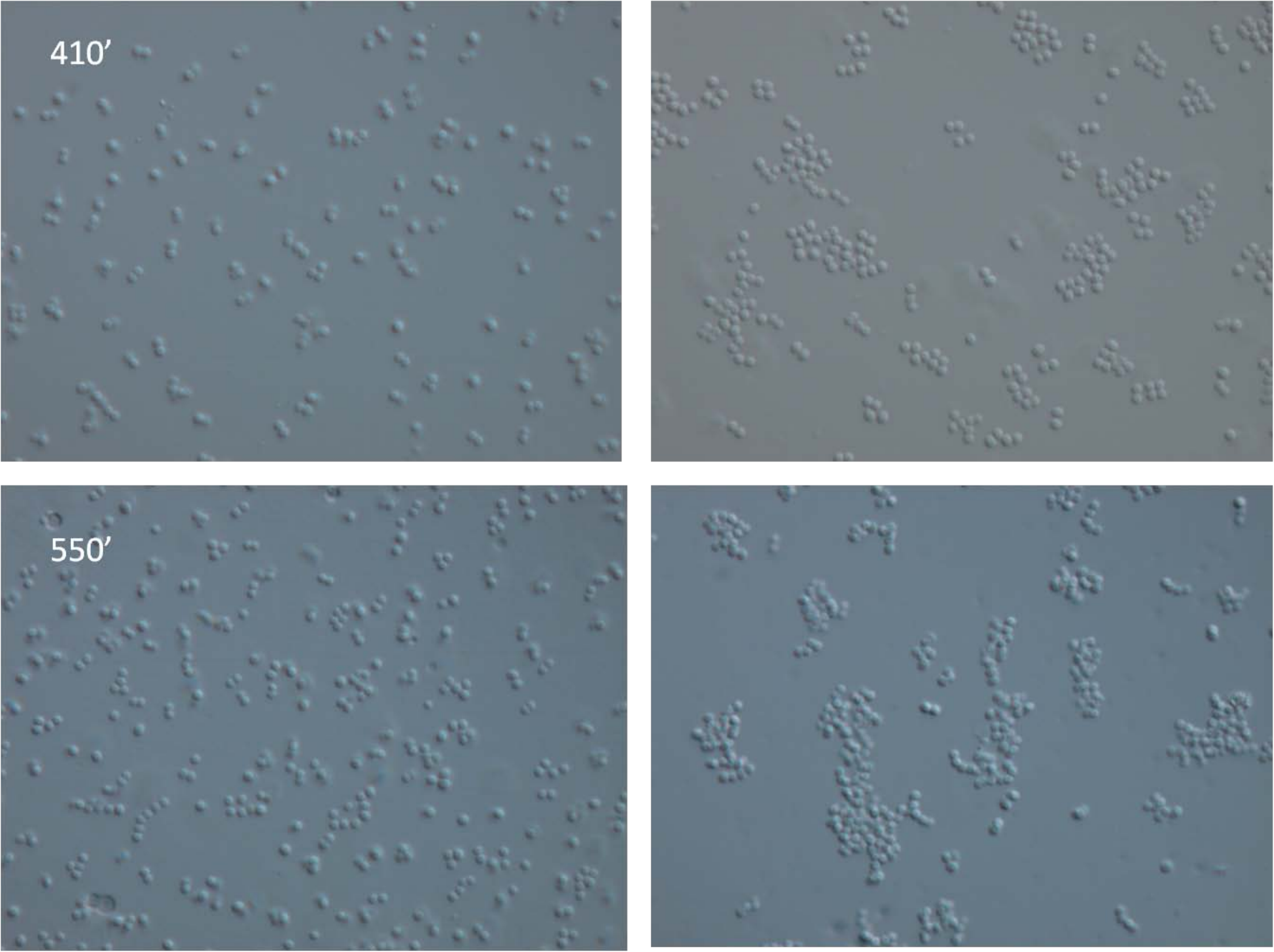}\\
\end{tabular}
\caption{Optical microscopy images obtained during growth of \textit{S. aureus} cultures -- strain COL (left panel) and strain RUSAL9 (right panel) at, respectively, 250, 350, 410 and 550 min (representative images). 10 photos were taken randomly from each aliquot.
}
\label{fig2}
\end{center}
\end{figure}

The deficient cell separation of mutant strain RUSAL9 is reflected in the optical microscopy images, as shown in Figure \ref{fig2}: the mutant always presents larger aggregates.

\begin{figure}[htp]
\begin{center}
\includegraphics[scale=0.45]{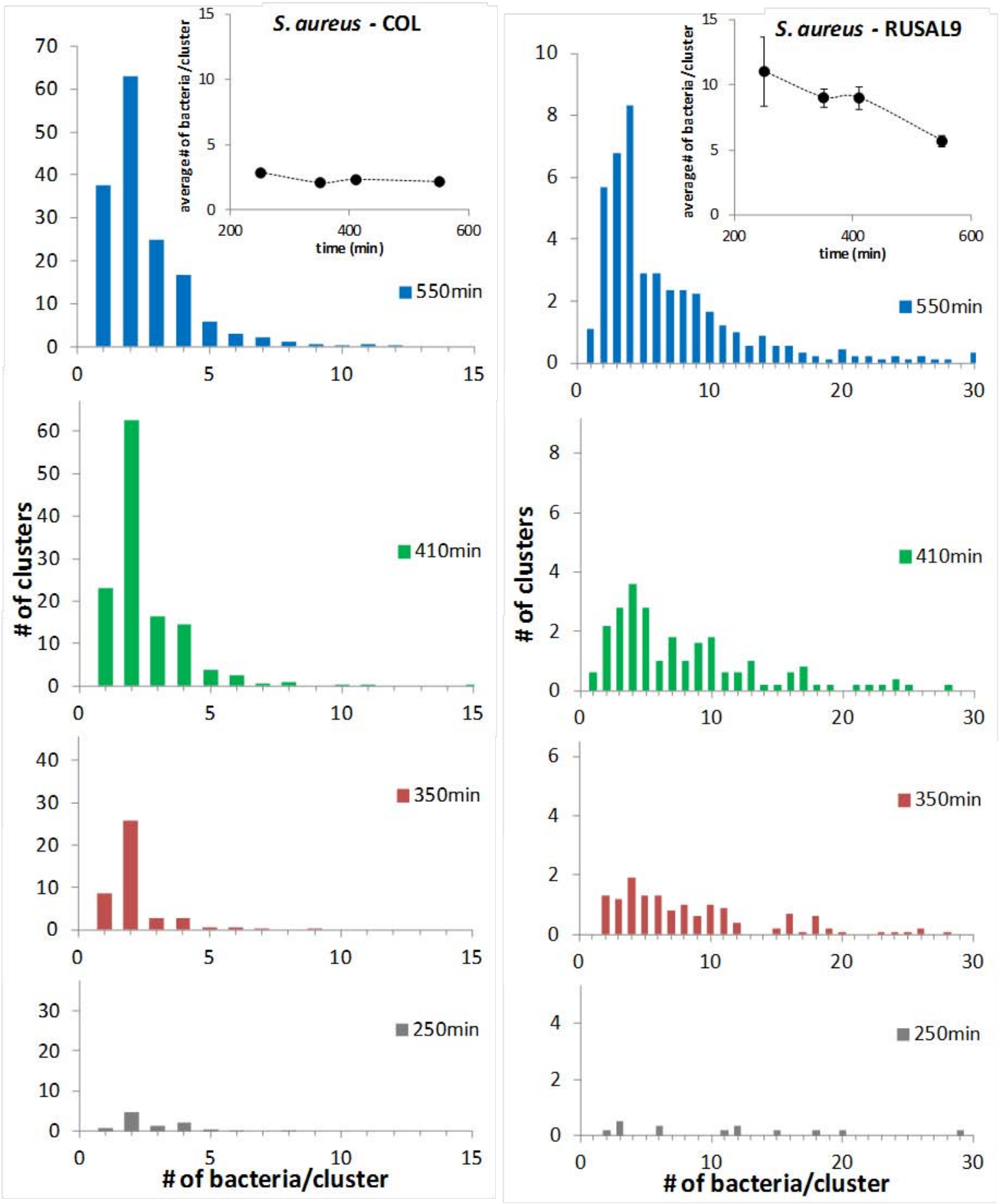}
\caption{(Colour online) Cell distribution during growth of \textit{S. aureus} cultures -- strain COL (left panel) and strain RUSAL9 (right panel), histograms: the number (\#) of cluster vs the number (\#) of in-plane cells per cluster  (counting only the cells shown in the plane of the image) at, respectively, 250 (grey), 350 (red), 410 (green) and 550 (blue) min, obtained from image analysis considering an average over 10 images; inserts: the average number (\#) of in-plane cells per cluster \textit{vs} time; error bars correspond to the standard deviation of the mean.}
\label{fig3}
\end{center}
\end{figure}

To characterize in detail the aggregates formed by the two different strains, the cell distribution during growth was determined (see Figure \ref{fig3}). From an image analysis considering an average over 10 random images, aggregation histograms -- the number of clusters \textit{vs} the number of cells per cluster (only cells shown in the plane of the image were considered) -- and the average number of cells per cluster were plotted, for each strain and at each specific moment of growth.

\textit{S. aureus} strain COL presents sharp asymmetric histograms, with a maximum value of 2 in-plane cells per cluster, for each moment of growth. The number of clusters increases with time but the overall histogram shape remains unchanged, spanning from 1 to approximately 15 in-plane cells per cluster.

\textit{S. aureus} strain RUSAL9 presents broader histograms, with larger number of cells per cluster at every stage of growth. The shape of the histograms does not however remain constant, and smaller clusters become more frequent as the late phase of growth is approached. This may explain the decrease of the average number of in-plane cells per cluster with time, that falls from 11 at 250 min, to 5 at 550 min.

\subsubsection{Viscosity growth curve}

\begin{figure}[htp]
\begin{center}
\includegraphics[scale=0.3]{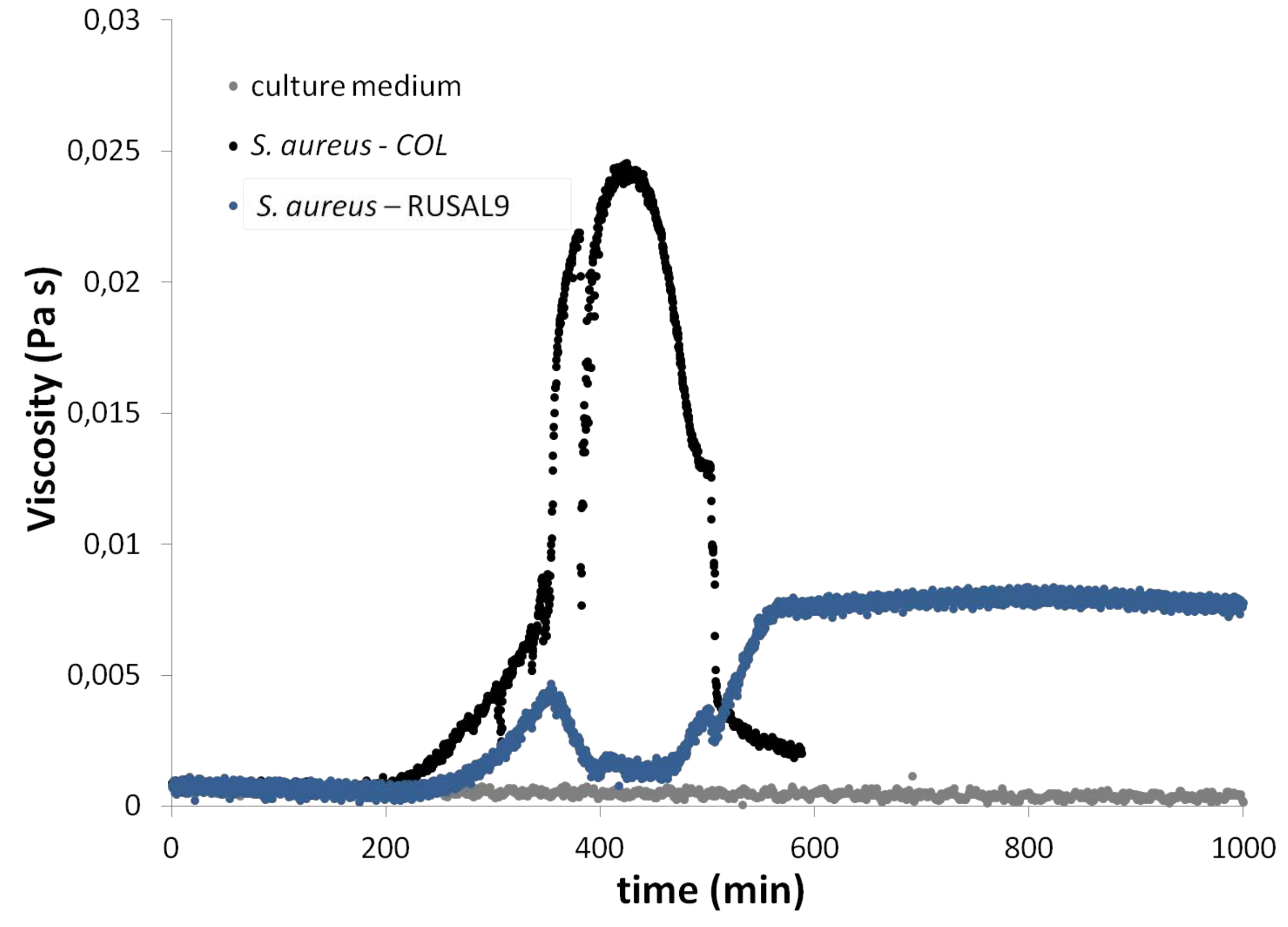}
\caption{(Colour online) \textit{S. aureus} cultures steady-state shear growth curves, $\eta(t)$ measured at a constant shear rate of  10s$^{-1}$ -- strain COL (black), strain RUSAL9 (blue) and culture medium (gray) (representative curves). All measurements performed at 37 $^\circ$C.}
\label{fig4}
\end{center}
\end{figure}

From the distinct time-dependent aggregation patterns described above, we may expect different mechanical properties for the two different strains. In particular, the viscosity growth curves shown in Figure \ref{fig4} exhibit very rich behaviours, with no counterpart in the optical characterization made so far.

The general behaviour of the viscosity growth curves show, for both strains, three distinct phases, consistent with the time intervals previously defined for the \textit{lag}, \textit{exponential} and \textit{late phases}.

In the first 300 minutes of growth, corresponding to the \textit{lag phase}, bacteria are adapting to the new environmental growth conditions, showing a characteristic slow division rate (see $\text{OD}_{620\text{nm}}$, Figure \ref{fig1}). In accordance, the viscosity of both strains shows a slow and constant increment with time.  However, in subsequent phases, significant differences can be pointed out.

In the \textit{exponential phase} (300 - 450 min), the $\text{OD}_{620\text{nm}}$ strongly increases in both strains.
In strain COL, this was accompanied by a dramatic increase in the viscosity by a factor of 30 ($\sim 10\times$ viscosity increase in this interval). However, the viscosity increase in this period is not monotonic, but exhibits several drops and recoveries. In particular, at around 400 min, as the viscosity was reaching its maximum value, there was a very large sudden decrease followed by an immediate recovery in a $\sim 15$ min time interval. In the case of strain RUSAL9, the viscosity in the \textit{exponential phase} also starts to increase (up to $\sim 5\times$ its initial value), but, at a certain threshold (360 min, see Figure \ref{fig4}), an abrupt change in its slope occurs, and the viscosity returns again to a value close to the initial one. No recoveries are observed until the end of this phase.

At 450 - 550 min, depending whether we consider COL or RUSAL9, the cfus/ml starts to stabilize, most probably due to nutrient depletion and accumulation of secondary metabolites which inhibit cell division. This corresponds to the beginning of the \textit{late phase}. For this time window, the behaviour of the viscosity growth curves was dramatically different. In the case of strain COL, at approximately 450 min the viscosity reached its maximum and decreased rapidly to a value close to its initial one. Instead, in the case of strain RUSAL9, the viscosity increased again, reaching, at an abrupt threshold (560 min), a plateau with a maximum value
$\sim 9\times$ larger than the initial one.

To understand these results, we resort to a microscopic description of the bacterial cells, as they divide, disperse and fill the system. This has been already initiated in our recent work \cite{Portela2013} and allowed us to interpret the viscosity growth curve of strain COL.

At initial states, strain COL culture may be considered as a colloid of small spheres (of radius $\sim 0,5$ $\mu$m) dispersed in liquid medium. This type of bacterial species is known to produce, during the \textit{exponential phase} of growth, cell-surface proteins called adhesins, which promote cell-cell contact (or between bacterial cells and other cells or physical substrates) \cite{Tompkins1992,Voyich2005}. For low cell densities (at the beginning of growth), we may find them in small clusters of 5-15 units (see Figures \ref{fig2} and \ref{fig3}). During bacterial cell division, these aggregates grow and become often unstable, and small clusters separate and disperse in the medium. The reason for this separation may be of entropic nature, or an active strategy to explore richer regions in nutrients. As the density increases, the cell aggregates start to establish new contacts and form frequently a web or cellular structure. Although these structures are not clear in our optical images (Figure \ref{fig2}), they most likely exist and were observed in similar systems \cite{Salek2012,Haaber2012}. As the percolation threshold is overcome, a rapid increase of the viscosity is expected, as observed in Figure \ref{fig4}, in the \textit{exponential phase}. The complex percolated structures which are able to develop during shear, may also explain the sudden viscosity drops and recoveries observed in very small periods of time during this period of global viscosity increase of the growth curves, and culminating in the intense asymptotic viscosity decrease and recovery near $\sim 400$ min (see Figure \ref{fig4}). It is expected that these relatively low density percolated structures which usually block or jam the motion, unblock and allow in small periods of time a release in the stress applied to the plates, which corresponds to a viscosity decrease. As the density of bacterial cells increases, unjamming transitions become more intense, until eventually the system reaches a high enough density, for which no more stress release is expected. The system tends then to stabilize its viscosity. Indeed, the viscosity increase ceases around $\sim 450$ min, as the evolution of cfus/ml (see Figure \ref{fig1} and \ref{fig4}) stabilizes in its highest value, and where the \textit{late phase} begins. From this moment, although the number of cfus/ml remains constant (indicating an unaltered population viability), the viscosity dramatically decreases close to its initial value. A change in bacterial physiology is probably the cause of this striking behaviour. In the \textit{late phase}, a decrease in the production of adhesins occurs \cite{Voyich2005}. Without being able to adhere, bacterial cells do not resist the flow, and viscosity decreases close to its original value. Although the assays were continued further for several hours, the viscosity remains unchanged. We hypothesize that, for this high cell density, the quorum-sensing mechanisms are repressing the production of adhesive factors and the population is no longer able to re-establish the percolated structure.

Due to the deficient daughter cell separation of strain RUSAL9, the irregular aggregates, formed at low densities, may easily reach several dozens of cells (see Figures \ref{fig2} and \ref{fig3}). Upon entering the \textit{exponential phase}, these aggregates start to be connected, and the viscosity begins to increase noticeably. No drops and recoveries are observed. In fact, as soon as the viscosity reaches a certain threshold (360 min), it recedes and does not recover again, until the end of the \textit{exponential phase}. Being composed of much larger and irregular aggregates than in the case of strain COL, the incipient percolated structure of strain RUSAL9 is easily undone by shear. With growth, the cell density increases, but the formation of a new percolated structure seems to be no longer possible, and the viscosity remains small and unchanged for a while (until $\sim 460$ min). Eventually, at higher bacterial density, the viscosity increases again reaching its maximum value. At a given (560 min) moment, which roughly coincides with the beginning of the \textit{late phase} of growth, the slope changes abruptly, and the viscosity reaches a plateau that
is maintained throughout the remaining time frame ($560-1000$ min). As the cells fail to separate upon division, at this stage we can not infer on the relative contribution of the bacterial adhesion properties. However, the observed final higher viscosity, when compared to the final viscosity of strain COL, must be directly related to the existence of much larger clusters, where the cells remain strongly ``glued'' to each other.

\subsection{Bacterial population dynamics dependence on external mechanical stimuli}

\subsubsection{Viscoelastic behaviour -- shear rate effect, stationary flow}

\begin{figure}[htp]
\begin{center}
\begin{tabular}{cc}
\includegraphics[scale=0.3]{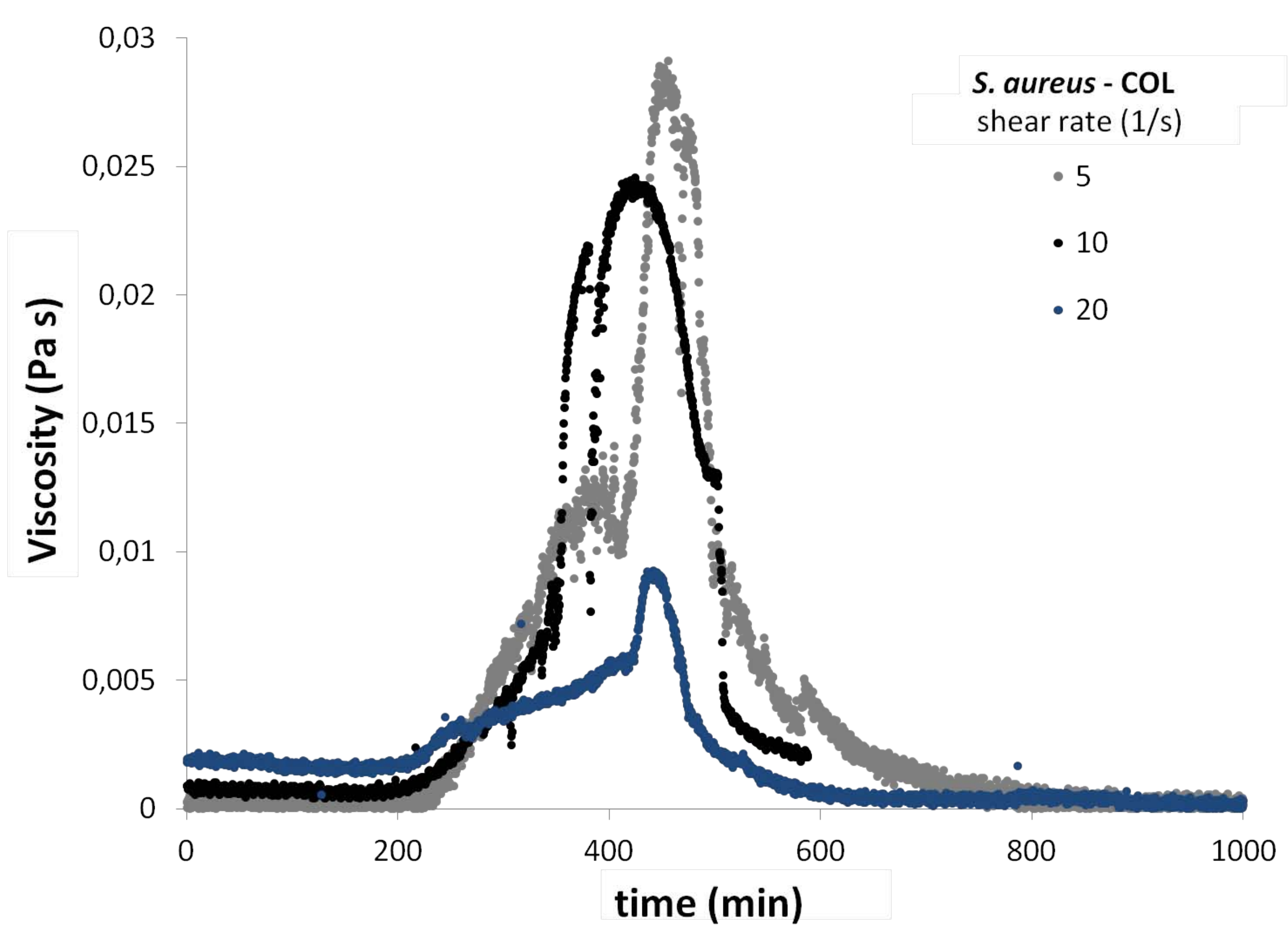}\\
\includegraphics[scale=0.3]{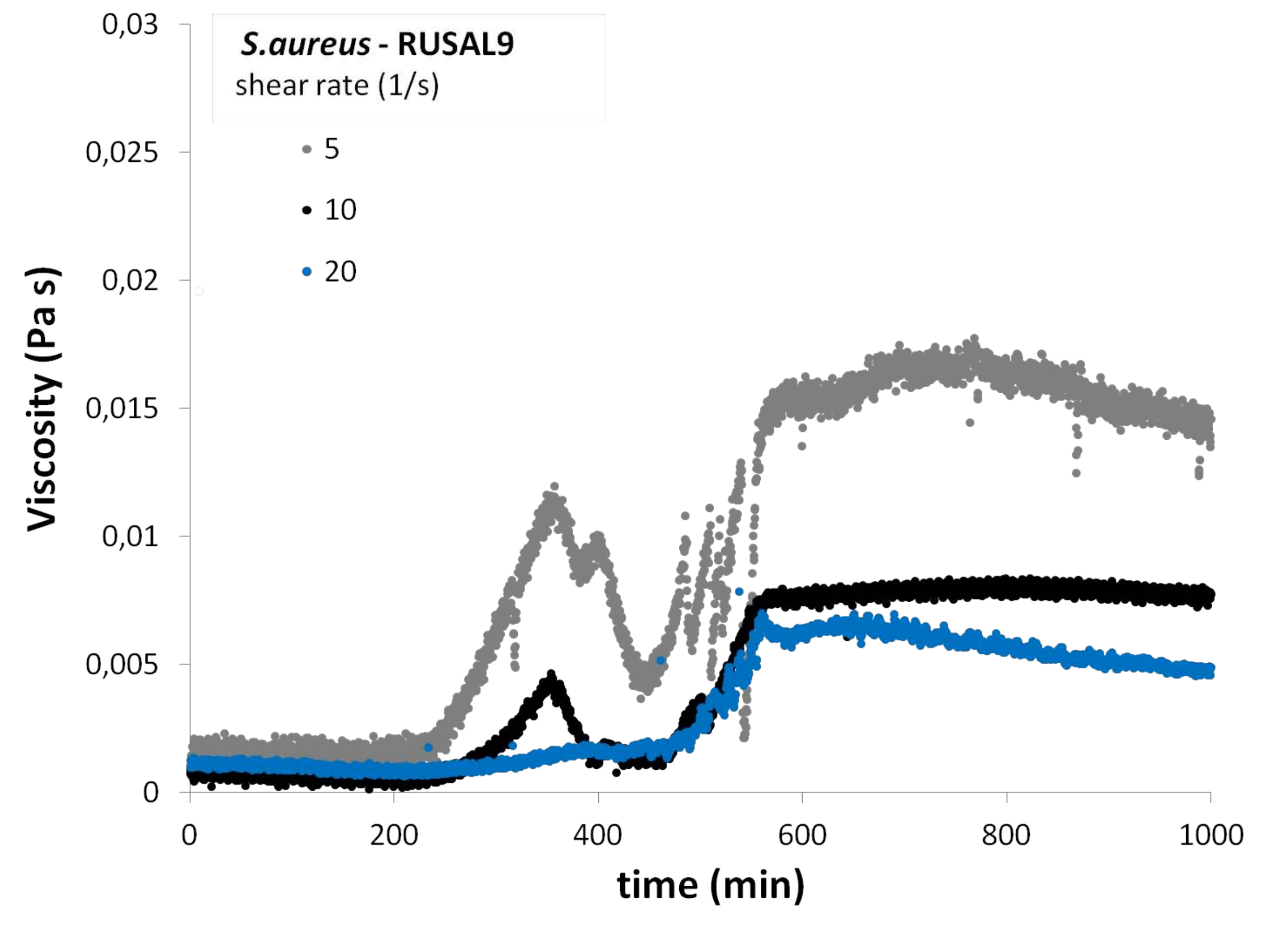}\\
\end{tabular}
\caption{(Color online) \textit{S. aureus} cultures -- a) strain COL and b) strain RUSAL9, steady-state shear growth curves, $\eta(t)$, measured at a constant shear rate of  5 (grey), 10 (black) and 20 (blue) s$^{-1}$; all measurements performed at 37 $^\circ$C.}
\label{fig5}
\end{center}
\end{figure}

To understand to what extent the behaviour of the bacterial culture (cell separation and aggregation properties) is affected by the intensity of shear, we measured the viscosity growth curve also for 5 and 20 s$^{-1}$. The results obtained for strain COL and strain RUSAL9 are represented in Figure \ref{fig5} a) and b), respectively.

For each strain, the general shape of the viscosity growth curves remains the same. The three different growth phases may still be distinguished and occur in the same time intervals. However, in general, higher shear rates lead to smaller viscosity values. This may be explained by our microscopic model: higher shear rates will difficult the formation of percolated structures of higher complexity, responsible for the larger values of viscosity. In the case of strain COL, when the production of adhesins is inhibited (\textit{late phase}), the decrease in viscosity also occurs faster as the shear rate increases. In the case of strain RUSAL9, the maximum viscosity attained in the \textit{late phase} decreases for higher shear rates. Interestingly, in the beginning of the \textit{exponential phase}, this strain shows a few drops and recoveries for the lower shear rate, whereas almost any significant increase of the viscosity occurs for the higher shear rate at this interval (300-400 min).
In this last case, the viscosity remains small almost until the end of the \textit{exponential phase}.

\subsubsection{Viscoelastic behaviour -- oscillatory flow}

To characterize the dependency of the elastic and the viscous moduli, $G'$ and $G''$, on the angular frequency $\omega$, oscillatory flow measurements were performed to aliquots of \textit{S. aureus} cultures, at specific moments of growth. The aliquots were sampled from bacterial culture grown in optimal laboratory conditions
(37 $^\circ$C).

\begin{figure}[htp]
\begin{center}
\begin{tabular}{cc}
\includegraphics[scale=0.3]{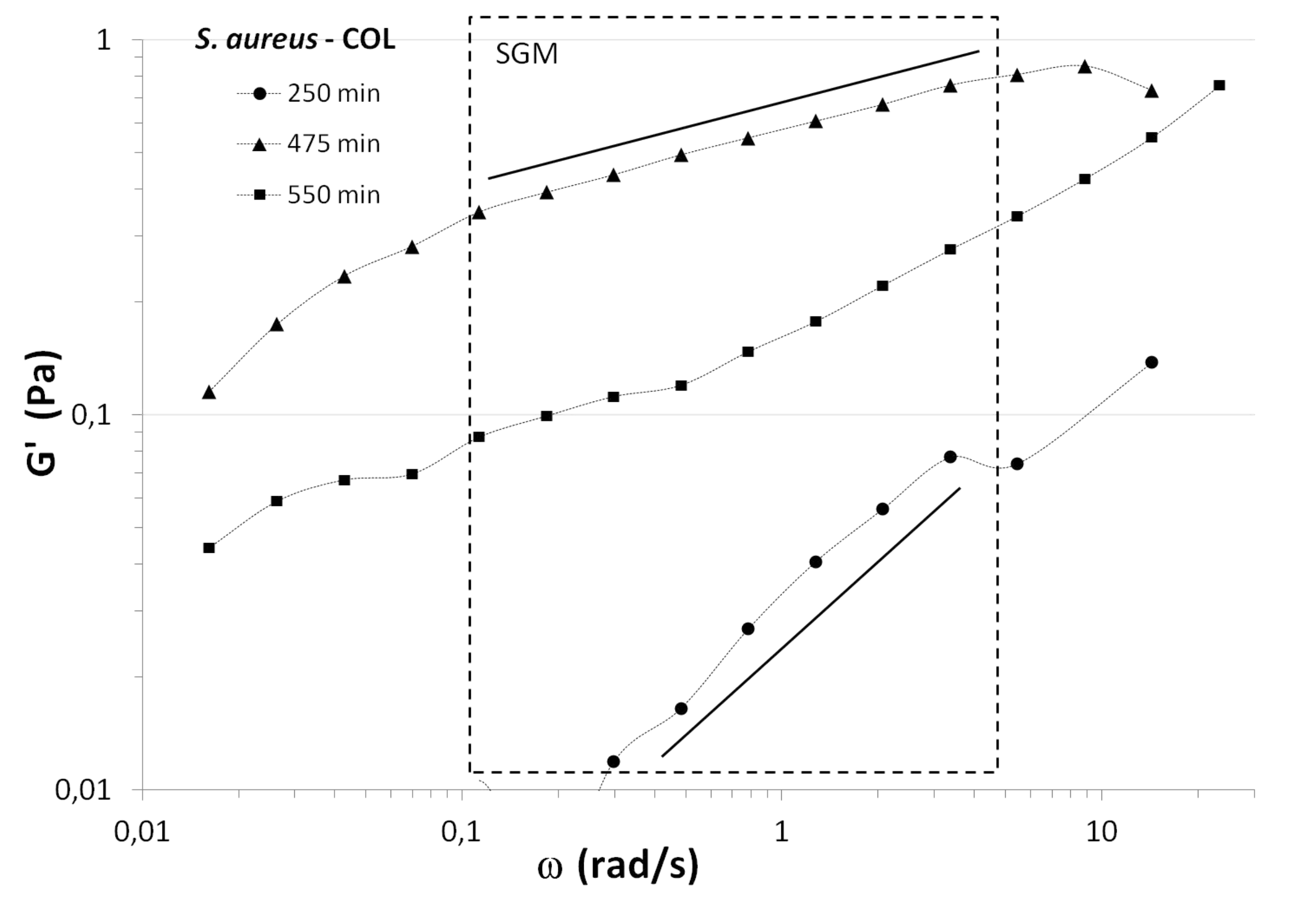}\\
\includegraphics[scale=0.3]{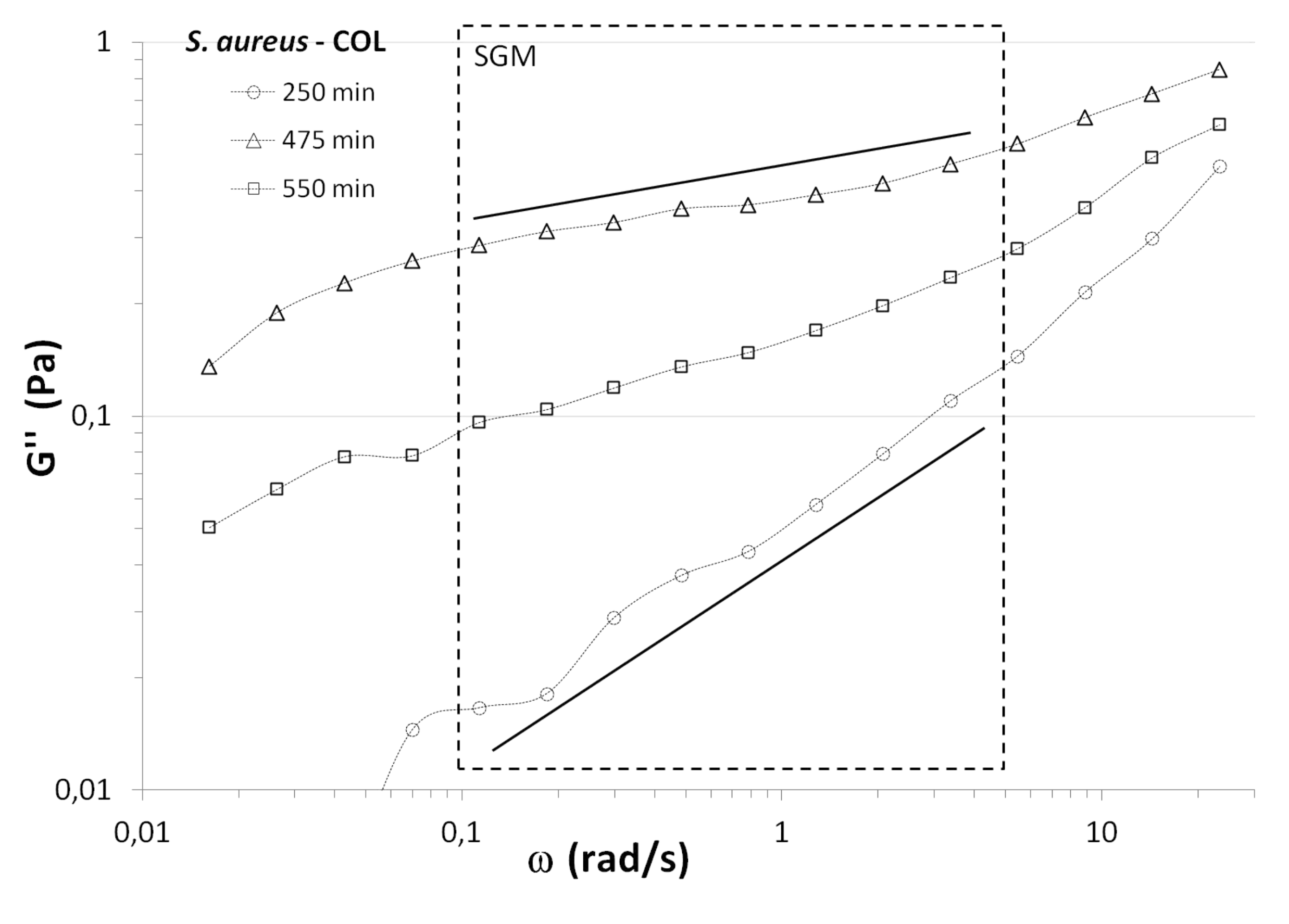}\\
\end{tabular}
\caption{Oscillatory shear flow experiments of the \textit{S. aureus} -- strain COL; a) elastic modulus $G'$ and b) viscous modulus $G''$ dependence on the angular frequency $\omega$, for culture aliquots with 250, 475 and 550 min of growth time; dashed box delimits the same power-law dependence of $G'$ and $G''$ with $\omega$ as defined by the Soft Glassy Materials model, solid lines are guides to the eye, corresponding to specific power laws; all measurements were performed at 20 $^\circ$C (to minimize bacteria growth during tests).}
\label{fig6}
\end{center}
\end{figure}

\begin{figure}[htp]
\begin{center}
\includegraphics[scale=0.3]{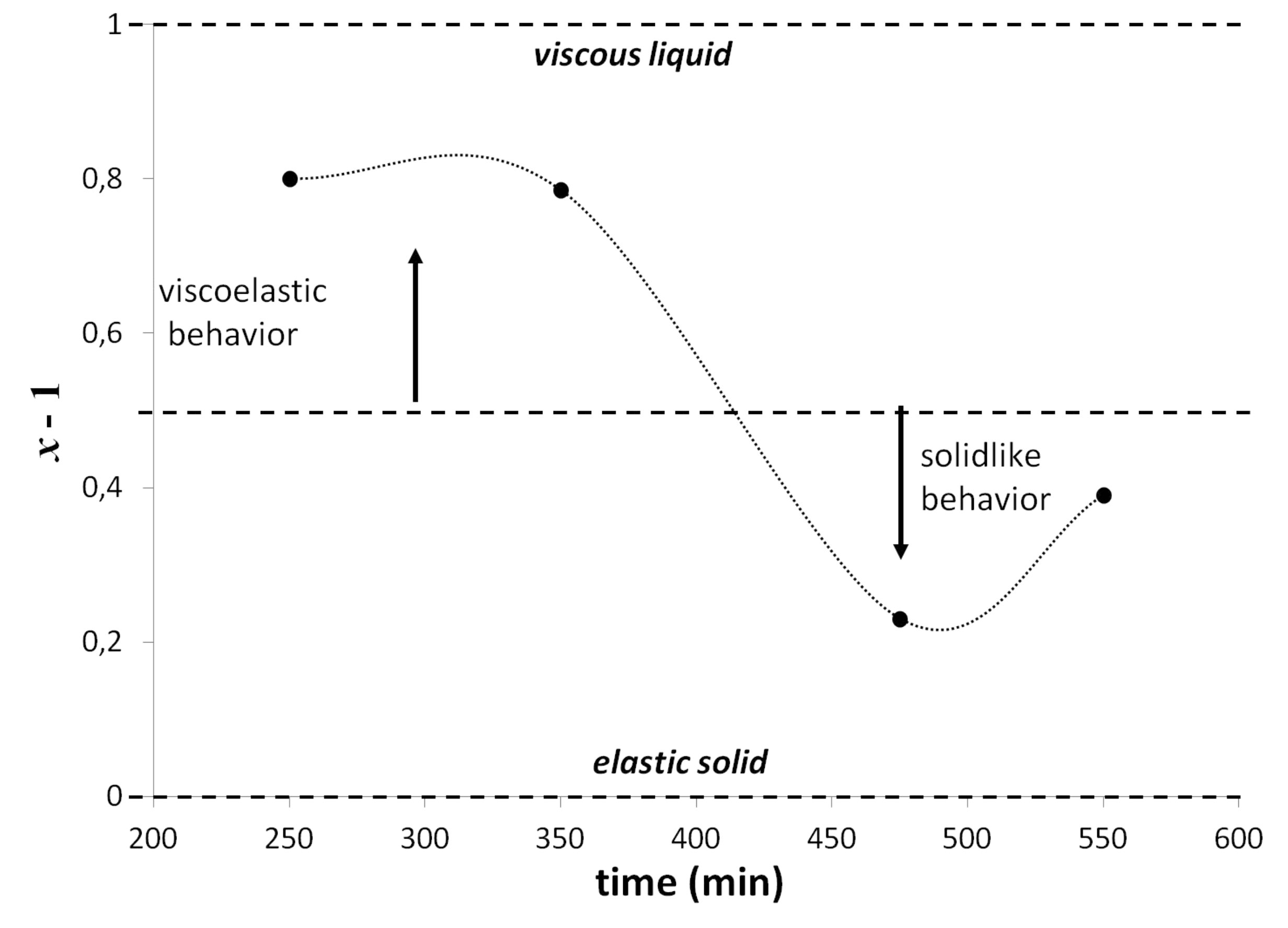}
\caption{Power law exponent $(x - 1)$ \textit{vs} time (solid dots), obtained from the fits to the previous oscillatory shear flow $G'$ and $G''$ measurements (Figure \ref{fig5}); the pointed line is a guide to the eye and the dashed lines correspond to the perfect solidlike, viscoelastic and fluidlike behaviours.}
\label{fig7}
\end{center}
\end{figure}

The results obtained for strain COL are represented in Figure \ref{fig6}.
As previously noted \cite{Portela2013}, for each moment of growth, the elastic and viscous moduli, $G'$ and $G''$, increase with $\omega$, following approximately the same weak power-law, at least in some range of angular frequencies indicated by the dashed rectangular boxes. This behaviour is well framed in the context of the Soft Glassy Material (SGM) model \cite{Sollich1997,Sollich1998}, where disorder and metastability are essential features. The analogy has been applied to other living cell systems \cite{Fabry2001,Rogers2008,Wilhelm2008}. In SGM model, the system exhibits an energy landscape with multiple minima, with high energy barriers (higher than thermal energy) and a distribution of yield energies. These energies may be overcome by imposing higher angular frequencies. The hopping between minima is activated by an effective noise temperature $x$, which is related to how jammed the system is. This model system predicts weak power law dependencies $G'\sim\omega^{x-1}$ and $G''\sim\omega^{x-1}$. When $x=1$ the system behaves as a perfect elastic body (solidlike) and when $x>1$ the system can flow and becomes disordered (fluidlike when $x=2$). The exponents calculated for each aliquot show a time dependency that is consistent with the various stages of growth, see Figure \ref{fig7}. The results show a fluidlike behavior ($x=1.80$) for the early stages of growth, when the percolated structures possibly start to appear. At the peak associated with the 475 min aliquot, the system has an almost solidlike behaviour ($x=1.23$). Finally, and remarkably, the system becomes again more fluid. At this point, the cell density does not decrease, but their physiological activity has somehow changed.
Our explanation is based on the hypothesis that there is less adhesion between cells. The SGM model also predicts the approximate law $G''/G'
\approx\tan\pi (x-1)/2$ (see Equation (1) of \cite{Fabry2001}). In the region delimited by the power-law fits, it is possible to find a ratio $G''/G'\approx 2.4$ for the first stages of growth and a ratio $G''/G'\approx 0.6$ at 475 min. These results agree reasonably well with the prediction $G''/G'=3.1$ and $G''/G'=0.4$ when the power-law exponents are $x=1.80$ and $x=1.23$, respectively. In the final stages of growth, as the exponents approach $x=1.5$ (see Figure \ref{fig7}), the elastic and viscous moduli become almost identical, as expected.

\begin{figure}[htp]
\begin{center}
\begin{tabular}{cc}
\includegraphics[scale=0.3]{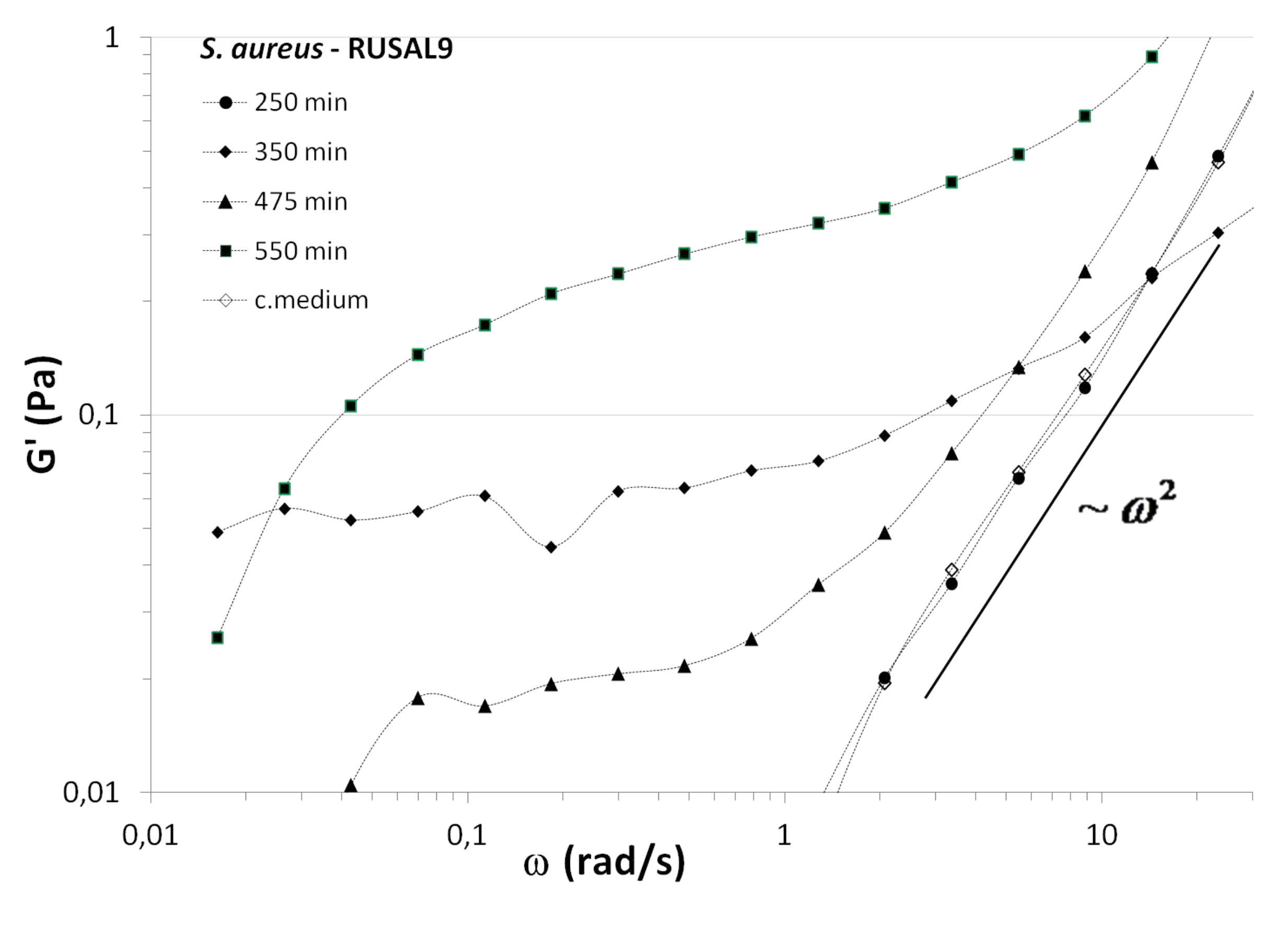}\\
\includegraphics[scale=0.3]{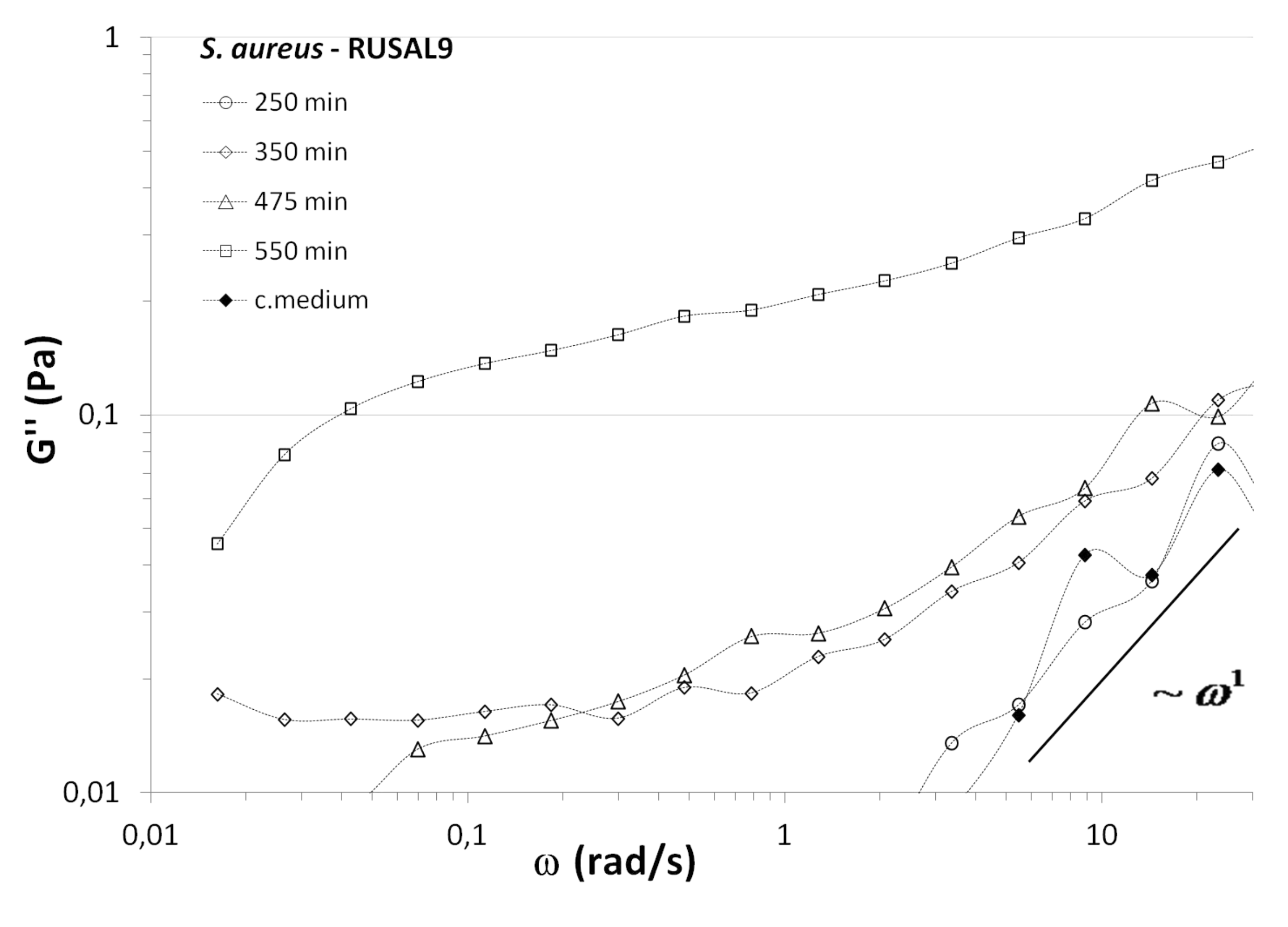}\\
\end{tabular}
\caption{Oscillatory shear flow experiments of \textit{S. aureus} -- strain RUSAL9; a) elastic modulus $G'$ and b) viscous modulus $G''$, dependence on the angular frequency $\omega$, for culture aliquots with 250, 350, 475 and 550 min of growth time and culture medium; solid lines are guides to the eye, corresponding to specific power laws; all measurements were performed at 20 $^\circ$C (to minimize bacteria growth during tests).}
\label{fig8}
\end{center}
\end{figure}

The same oscillatory flow measurements were applied to strain RUSAL9 aliquots and the results can be observed in Figure \ref{fig8}. In this case, $G'$ and $G''$ no longer follow the weak power law behaviours, but show a more complex dependence with $\omega$ denoting the presence of distinct relaxation processes. For lower angular frequencies, $G'$ and $G''$ essentially reflect the self-organized structures developed by the bacteria, according to the successive stages of growth (the culture medium shows $G'$,$G''<10^{-3}$ Pa). $G'$ increases strongly at 350 min (beginning of the \textit{exponential phase}), decreases to a lower value at 475 min (as it is observed for the viscosity in the steady shear flow, see Figure \ref{fig4}), and increases again at 550 min (as occurs for the viscosity in the \textit{late phase} of growth). $G''$ presents a similar behaviour. However, the curves obtained for 350 and 475 min remain similar. For higher angular frequencies, the influence of the culture medium emerges and we obtain, for almost all aliquots, the familiar Maxwell power law behaviours, $G'\sim\omega^2$ and $G''\sim\omega$.

\begin{figure}[htp]
\begin{center}
\includegraphics[scale=0.3]{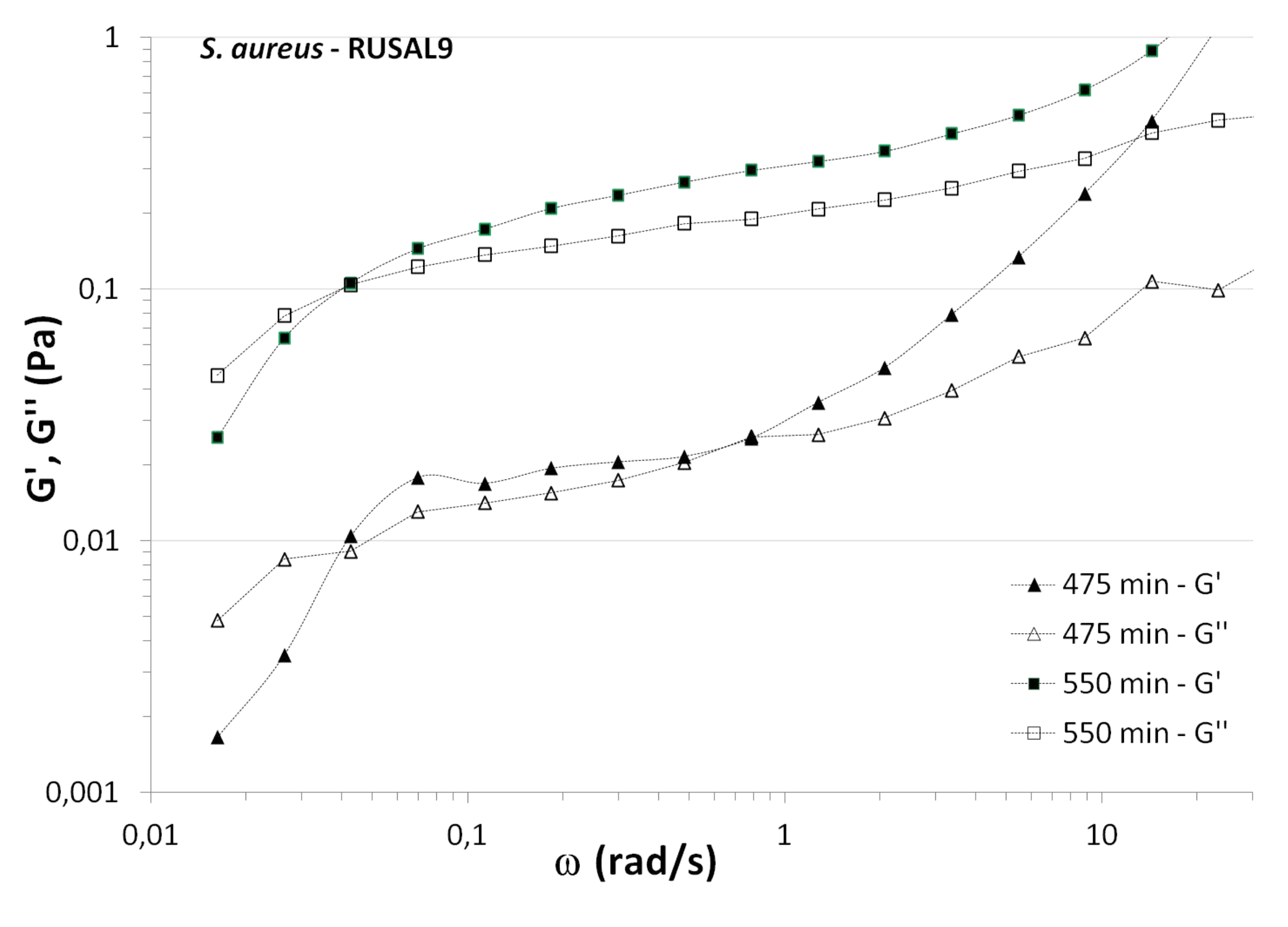}
\caption{Oscillatory shear flow experiments of \textit{S. aureus} -- strain RUSAL9, elastic modulus $G'$ and viscous modulus $G''$, dependence on the angular frequency $\omega$, for culture aliquots with 475 and 550 min of growth; all measurements were performed at 20 $^\circ$C (to minimize bacteria growth during tests).}
\label{fig9}
\end{center}
\end{figure}

It is noteworthy that our complex system will not have only one, but surely several relaxation modes. The larger relaxation modes, probably associated with the superstructures created by the bacteria, may be seen in Figure \ref{fig9}, in which we have represented both $G'$ and $G''$, for the last aliquots, at 475 and 550 min. In both cases, and at lower angular frequencies, $G'$ crosses $G''$ at $\omega\approx 0.04$ rad/s, an indication of a relaxation time of $\tau\approx 150$ s. In this figure, a second relaxation time, probably associated with the large molecules that compose the culture medium, seems to arise at much higher angular frequencies ($\omega>100$ rad/s).

To understand the behaviours shown in Figure \ref{fig9}, it is interesting to consider the generalized Maxwell's model with two modes of relaxation, described by the following equations:
\begin{eqnarray}
G'=G_1\frac{(\omega\tau_1)^2}{1+(\omega\tau_1)^2}+G_2\frac{(\omega\tau_2)^2}{1+(\omega\tau_2)^2}\\
G''=G_1\frac{\omega\tau_1}{1+(\omega\tau_1)^2}+G_2\frac{\omega\tau_2}{1+(\omega\tau_2)^2}
\end{eqnarray}

\begin{figure}[htp]
\begin{center}
\begin{tabular}{cc}
\includegraphics[scale=0.3]{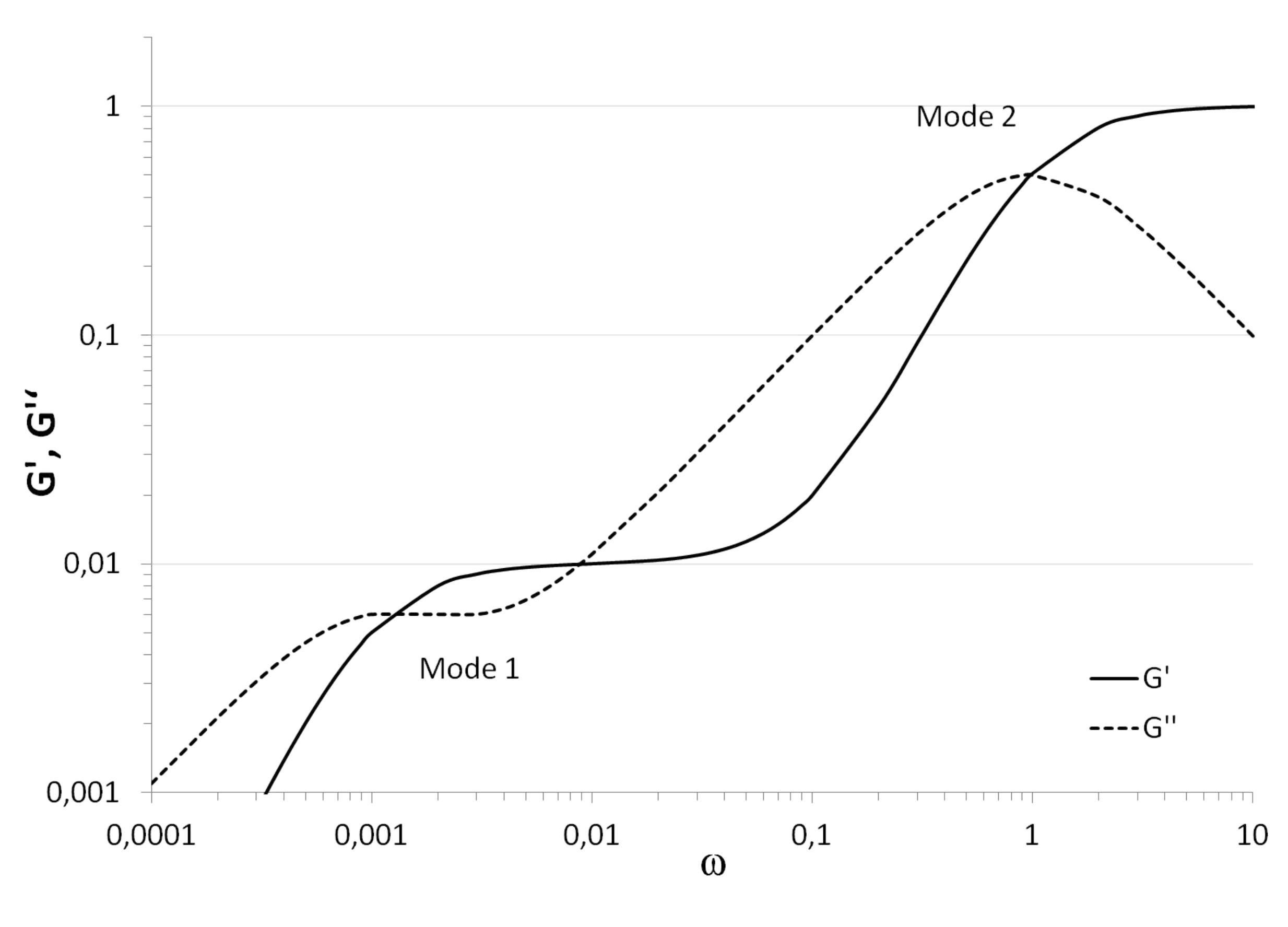}\\
\includegraphics[scale=0.3]{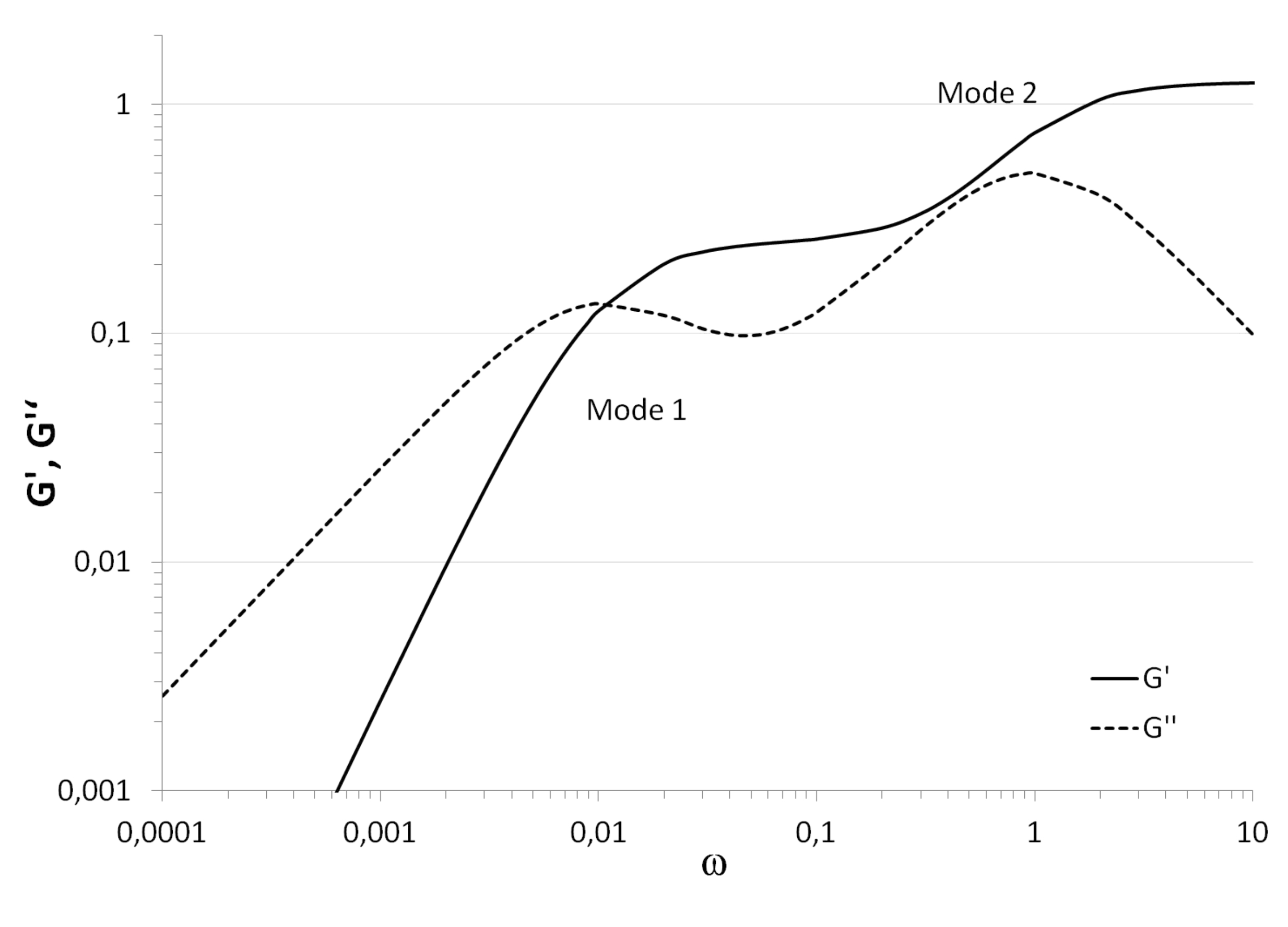}\\
\end{tabular}
\caption{Generalized Maxwell's model: elastic modulus $G'$ (full line) and viscous modulus $G''$(dashed line) as a function of the angular frequency $\omega$, in adimensional units, considering two different relaxation modes: a) $\tau_2=1000\tau_1$ and $G_2=0.01G_1$ and b) $\tau_2=100\tau_1$ and $G_2=0.25G_1$.}
\label{fig10}
\end{center}
\end{figure}

Two different scenarios are proposed in Figure \ref{fig10}. Figure \ref{fig10} a) represents the case of two very different relaxation times, $\tau_2=1000\tau_1$ and $G_2=0.01G_1$, which almost do not interfere with one another. These relaxation times are clearly detectable by the points of intersection of $G'$ and $G''$ lines (at $\omega=0.001$ and $\omega=1$, adimensional units). Before the points of intersection, we have $G'\sim \omega^2$ and $G''\sim\omega$ as verified by the experimental measurements at large values of $\omega$ (see Figure \ref{fig9}). In Figure \ref{fig10} b), the relaxation times are closer, $\tau_2=100\tau_1$, $G_2=0.25G_1$, and the relaxation modes start to interfere. For intermediate and large angular frequencies, a complex behaviour is obtained, in this case without any intersection of $G'$ and $G''$. There is only one intersection, corresponding to the slowest relaxation mode (at $\omega=0.01$, adimensional units). This second scenario seems to give a good qualitative description of the viscoelastic behaviour presented in Figure \ref{fig9}, obtained for strain RUSAL9 at 475 and 550 min.

\section{Conclusions}
\label{sec4}

In this study, real-time and \textit{in situ} rheology was applied to two related strains of the human pathogen \textit{Staphylococcus aureus} -- strain COL and its mutant RUSAL9, during cell growth. As the density of bacteria in the medium increases, going through different growth stages,
cells may rearrange themselves in different aggregates, capable of strongly influencing their environment,
and leading to very different physical viscoelastic responses.
The strains investigated in our work were intentionally chosen because they present distinct time-dependent aggregation patterns,
although maintaining the same genetic background.
Strain COL forms smaller clusters, while strain RUSAL9, presenting deficient cell separation, forms irregular larger aggregates.

The general behaviour of the viscosity growth curves show, for both strains, three distinct phases,
timely consistent with the \textit{lag}, \textit{exponential} and \textit{late phases}, also present in optical density and colony forming units measurements.
However, the richness observed in the viscosity growth curves, has no counterpart in the common microbiologic measurements.
Furthermore, the differences found in the viscosity profiles allows one to unequivocally distinguish the two strains here considered.
These complex viscoelastic behaviours are a consequence of the coupled contribution of two factors.
In the case of strain COL, the two factors are the cell density continuous increase and its changing cell-cell adhesion determinants, which production is most probably controlled by quorum-sensing mechanisms. In the case of strain RUSAL9, in addition to the cell density continuous increase, the second contribution factor is the impairment of the cell separation process. For this strain, the cell-cell interactions are mainly mediated by the deep share of cell wall material between sister cells at the septum. In this context, the native adhesion factors, characteristic of \textit{S. aureus}, would only have a secondary role.

To investigate how the bacterial population dynamics depends on the environmental stimuli, we followed bacterial growth under different shear flow conditions.
In stationary shear flow, the viscosity growth curves remained qualitatively the same. However, in general, higher shear rates led to smaller viscosity values.
The viscous and elastic moduli of strain COL, obtained for oscillatory shear,
exhibit power-law behaviours whose exponent are dependent on the bacteria growth stage, with a remarkable agreement with the Soft Glassy Material model.
The viscous and elastic moduli of strain RUSAL9 have complex behaviours, but in some cases consistent with a generalized Maxwell model with two relaxation modes.
These relaxation modes emerge from structures with two different time and length scales,
associated with the large molecules of the medium and the self-organized structures of bacteria.

As a technique used for the characterization of bacterial populations, rheology will only provide information on bulk properties of the population and not on
the nature of the bacterial behaviour, namely which substances are produced by the cells and in which quantity.
However, it is clear that the viscosity growth curves reflect the different aggregation properties of these two strains, and moreover, the progress of different physiological states (division rate/adhesion profile) during growth.
Standard classical rheology measurements emerge in this context as an innovative approach that complements the conventional visualization of macrobiological systems.
The ability to differentiate bacterial physiology profiles
opens new perspectives on the understanding of how environmental changes will impact on cell viability and population self-organized structures.

\section*{Acknowledgements}

Strain COL and Strain RUSAL9 were a kind gift from H. de Lencastre and A. Tomasz.
We acknowledge the support from FCT (Portugal) through Grant No. PEst-C/CTM/LA0025/2011 (CENIMAT/I3N),
PEst-OE/BIA/UI0457/2011 (CREM), and through Project PTDC/BIA/MIC/101375/2008 (awarded to RGS).

\bibliography{COLvsRUSAL9}

\end{document}